\documentclass{revtex4}

\pdfoutput=1 

\usepackage{graphicx,epstopdf,float}
\usepackage{bbm, bm, mathbbol, amsmath, longtable, amssymb,bbold}
\usepackage{color}
\usepackage{braket}

\newcommand{\beq}{\begin{equation}}
\newcommand{\eeq}{\end{equation}}

\begin{document}
\title{
High Efficiency Raman Memory by Suppressing Radiation Trapping}

\author{S.~E.~Thomas$^{1,2,*}$, J.~H.~D.~Munns$^{1,2}$, K.~T.~Kaczmarek$^1$, C.~Qiu$^{1,3}$, B.~Brecht$^1$, A.~Feizpour$^1$, P.~M.~Ledingham$^1$, I.~A.~Walmsley$^1$, J.~Nunn$^1$ and D.~J.~Saunders$^1$}

\affiliation{$^{1}$Clarendon Laboratory, University of Oxford, Parks Road, Oxford OX1 3PU, UK\\
$^{2}$ QOLS, Blackett Laboratory, Imperial College London, London SW7 2BW, UK\\
$^{3}$Department of Physics, Quantum Institute for Light and Atoms, State Key Laboratory of Precision Spectroscopy, East China Normal University, Shanghai 200062, People's Republic of China\\
$^*$Corresponding author sarah.thomas@physics.ox.ac.uk
}

\date{\today}

\begin{abstract}

Raman interactions in alkali vapours are used in applications such as atomic clocks, optical signal processing, generation of squeezed light and Raman quantum memories for temporal multiplexing. To achieve a strong interaction the alkali ensemble needs both a large optical depth and a high level of spin-polarisation. We implement a technique known as quenching using a molecular buffer gas which allows near-perfect spin-polarisation of over $99.5\%$ in caesium vapour at high optical depths of up to $\sim  2 \times 10^5$; a factor of 4 higher than can be achieved without quenching. We use this system to explore efficient light storage with high gain in a GHz bandwidth Raman memory.

\end{abstract}

\maketitle

\section{Introduction} 

The strong spin-orbit coupling in alkali vapours enables a broadband optical interface for spin coherence via Raman scattering. Optically dense alkali vapours can serve as a frequency reference for atomic clocks and magnetometry~\cite{Ludlow2013,Magnetometry}, a buffer in optical signal processing~\cite{Tucker2005}, and in the quantum domain, provide a source of squeezed light via four-wave mixing~\cite{Boyer2009,Lett2008} or a medium for storing and synchronising photons, via the Raman- and DLCZ-type quantum memory protocols~\cite{Lvovsky2009, DLCZ,Lukin2007, Novikova2008,Novikova2008a,Bucher2011,Hosseini2011,Namazi2015, Michelberger2015,Saunders2015}. In each of these examples, the desired light-matter coupling is a collective effect, where the coupling strength scales favourably with the optical depth $d$, which is proportional to the atomic density.

Very high atomic densities can be achieved without complex atom trapping by heating a vapour cell, but in vapour cell systems the ability to spin-polarise the ensemble by optical pumping~\cite{Happer1972} is hampered by radiation trapping at high densities~\cite{RadiationTrapping1998}. This problem is greatly mitigated in hollow waveguides, where there is only high optical depth in one dimension, but here high densities are challenging due to surface adsorption~\cite{Kaczmarek2015}. Here we show that introducing a molecular buffer gas into a vapour cell suppresses radiation trapping via collisional quenching and enables high quality spin-polarisation even at high temperatures. We use this system to demonstrate efficient light storage with very high and controllable gain, and these results demonstrate a route towards high efficiency storage of non-classical light.

\section{Optical Pumping in alkali vapours}

\begin{figure}
\begin{center}
\includegraphics[width=.5\linewidth]{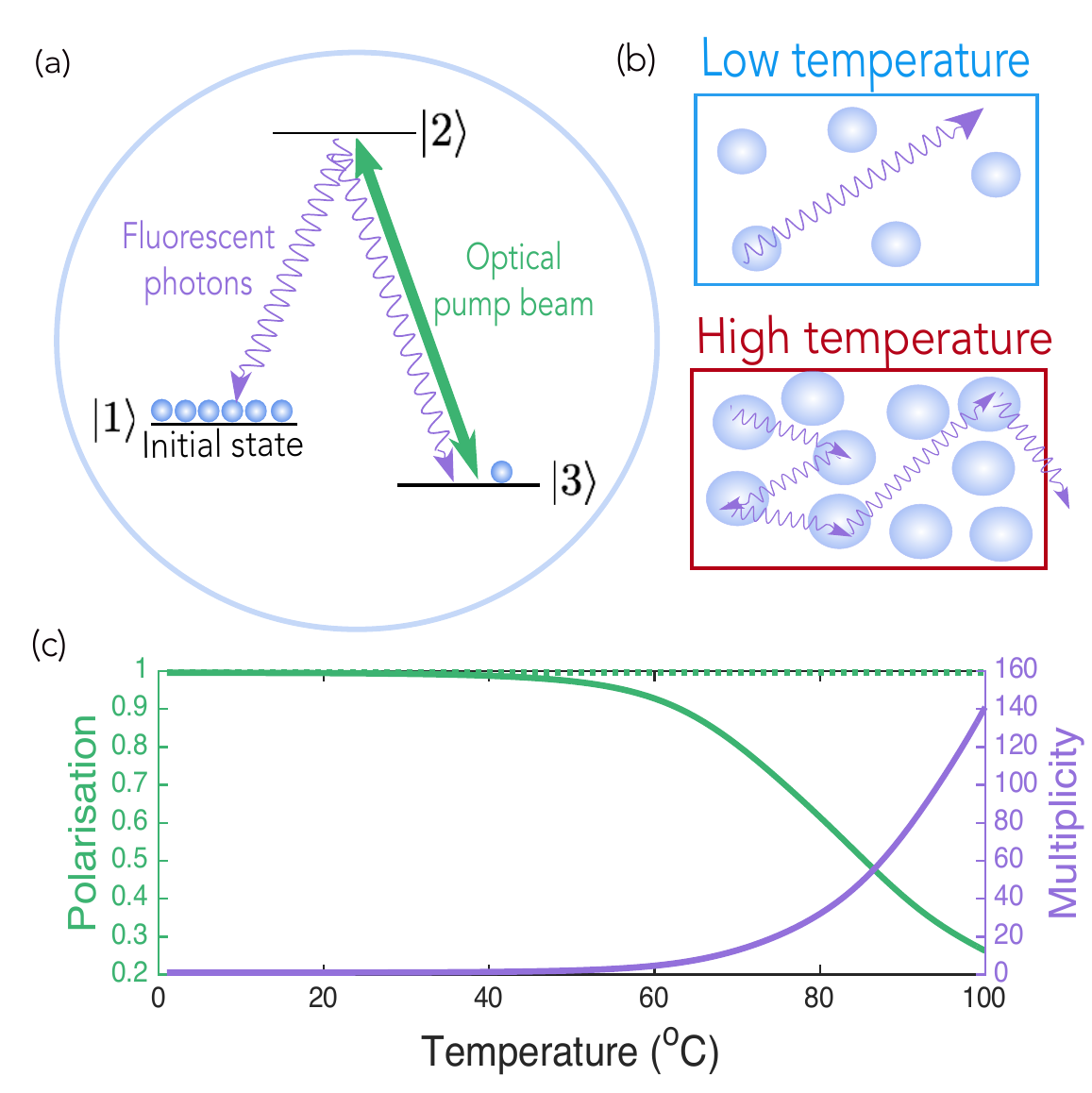}
\end{center}
\caption{(a) Optical pumping in a $\Lambda$-energy level system. (b) Radiation trapping. At higher temperatures, fluorescent photons are absorbed and emitted multiple times within the vapour cell, which depolarises atoms. (c) Multiplicity factor (purple) and predicted spin-polarisation (green) for caesium atoms in 20 Torr of neon buffer gas, using equations given in~\cite{Rosenberry2007}. The dashed green line is the predicted spin-polarisation if there is no radiation trapping.}
\label{figure1} 
\end{figure}

To enable strong Raman interactions, a high-density atomic ensemble needs to be prepared in a single ground state via optical pumping~\cite{Happer1972}. To initialise the ensemble in the ground state $\ket{1}$ (defined in Fig. \ref{figure1}(a)), a strong pump beam resonant with the $\ket{3} \leftrightarrow \ket{2}$ transition illuminates the ensemble. This excites atoms out of state $\ket{3}$, which then decay via fluorescence back to the ground states. Atoms in $\ket{3}$ are continuously depleted and the atoms all accumulate in $\ket{1}$ after several cycles of excitation and decay.

However, when an atom decays down to one of the ground states by fluorescence, it emits a photon. Thus a background population of photons resonant with $\ket{1} \leftrightarrow \ket{2}$ transition is established. This radiation can transfer an electron from $\ket{1}$ to $\ket{2}$, which can then decay to $\ket{3}$ - the reverse of optical pumping. If the alkali vapour is optically thick these photons can be absorbed and re-emitted multiple times, and ``un-pump'' many atoms in the vapour (see Figure~\ref{figure1}(b)). This effect is known as radiation trapping and limits the spin-polarisation that can be achieved when the vapour has a sufficiently high optical depth in multiple dimensions~\cite{RadiationTrapping1998}. 
 
The effect that radiation trapping has on optical pumping of alkali vapours is characterised using the multiplicity factor $M$, which is the average number of times a photon is re-absorbed and emitted before it leaves the vapour. It can be shown that $M$ increases exponentially in the number density of the vapour, which itself is exponential in temperature, and in a cylindrical bulk vapour cell $M$ can exceed 100 for typical operational conditions~\cite{Rosenberry2007}, as shown in Figure~\ref{figure1}(c). \\

In alkali vapour systems, a buffer gas is often added to the vapour to inhibit diffusion of the alkali atoms, and the choice of the species and pressure of the buffer gas plays a vital role in many experiments. To ensure high spin-polarisation, it is crucial that the collisions between the alkali atoms and the buffer gas are spin-preserving and induce very low rates of population transfer between the ground states. The spin-relaxation rate of polarised alkali vapours in different buffer gases has been studied, and is found to be low in inert gases such as the noble gases and molecular nitrogen~\cite{Franz1974}. Another important consideration for the choice of buffer gas is the relative trade-off between the diffusion rate versus the induced pressure broadening. In this work, we choose a buffer gas which also enables a non-radiative energy transfer pathway from atoms in the excited state $\ket{2}$ to the buffer gas via collisional quenching.

Our group previously used 20 Torr of neon as a buffer gas for Raman memory experiments as neon is spin-preserving and provides an appropriate trade-off between atomic diffusion and pressure broadening for our previous quantum memory experiments~\cite{Reim2011,Michelberger2015}. Fig.~\ref{figure1}(c) shows the numerical values for the multiplicity as a function of temperature for caesium (Cs) vapour with this choice of buffer gas. We can see that the radiation trapping effect becomes significant for $T\gtrsim 70^\circ$C, and this dramatically reduces the spin-polarisation at high temperatures. Here we investigate optical pumping in Cs vapour in an alternative buffer gas. If we can reduce radiation trapping and achieve high spin-polarisation at high temperatures, then we can increase the Raman interaction strength by operating at higher optical depths.

To suppress radiation trapping we introduce a mechanism whereby atoms decay from the excited state without fluorescence resonant with the $\ket{1} \leftrightarrow \ket{2}$ transition. If a molecular buffer gas is introduced which has a mechanical degree of freedom close in energy to the atomic transition, alkali atoms in the excited state collide with the molecules and transfer their energy to the molecule, thereby decaying to the ground state without emitting resonant photons. This reduces the number of resonant fluorescent photons produced in the optical pumping process and therefore suppresses radiation trapping, in a process known as collisional quenching. Quenching by introduction of a molecular buffer gas is a common technique used to limit radiation trapping, for example in spin-exchange optical pumping~\cite{Lancor:2010}, but this is the first time, to our knowledge, that is has been applied in the context of light storage.

To investigate the effect of collisional quenching, we compare the spin-polarisation of Cs vapour with two different buffer gases: neon (Ne) and molecular nitrogen (N$_2$). The latter gas provides a pathway for quenching as vibrational transitions exist which are very close in energy to the optical transitions of Cs~\cite{Happer1972}.
%

We first measured the diffusion of Cs in the two buffer gases using the method described in~\cite{Parniak2013} (see Supplementary Material). The results are given in Table~\ref{tab_D0vals}, and we find that the diffusion of Cs in N$_2$ is slower than in Ne at the same buffer gas pressure, and therefore for the case of a Raman memory, it is a viable buffer gas to ensure sufficient Raman interaction times for light storage on timescales $\sim 1 \mu$s. The pressure broadening of the Cs transitions is twice as high in N$_2$ buffer gas compared to Ne (Table~\ref{tab_D0vals}) and hence to ensure a fair comparison we investigate spin-polarisation in 20 Torr of Ne and with 10 Torr of N$_2$. \\

\begin{table}
\begin{tabular}{ccccc}
\hline\hline
Species & & $D_0\,[\mathrm{cm}^2\mathrm{s}^{-1}]$ &&$\gamma\,[\mathrm{MHz/Torr}]$ \\ \hline
N$_2$ &&
$0.24\pm0.09$ && $19.18\pm0.06$ \\
Ne &&
 $0.35\pm0.05$ && $9.81\pm0.06$ \\ \hline\hline

\end{tabular}
\caption{\label{tab_D0vals} The diffusion constants and pressure broadening coefficients for the two buffer gases considered. The diffusion parameters are extracted experimentally using the method presented in~\citep{Parniak2013}, and the errors denote \mbox{$95\%$} confidence bounds.  The pressure broadening coefficients are at a temperature of \mbox{$313\,\mathrm{K}$} quoted from~\citep{Pitz2010}.
}

\end{table}

\begin{figure}
\begin{center}
\includegraphics[width=0.9\linewidth]{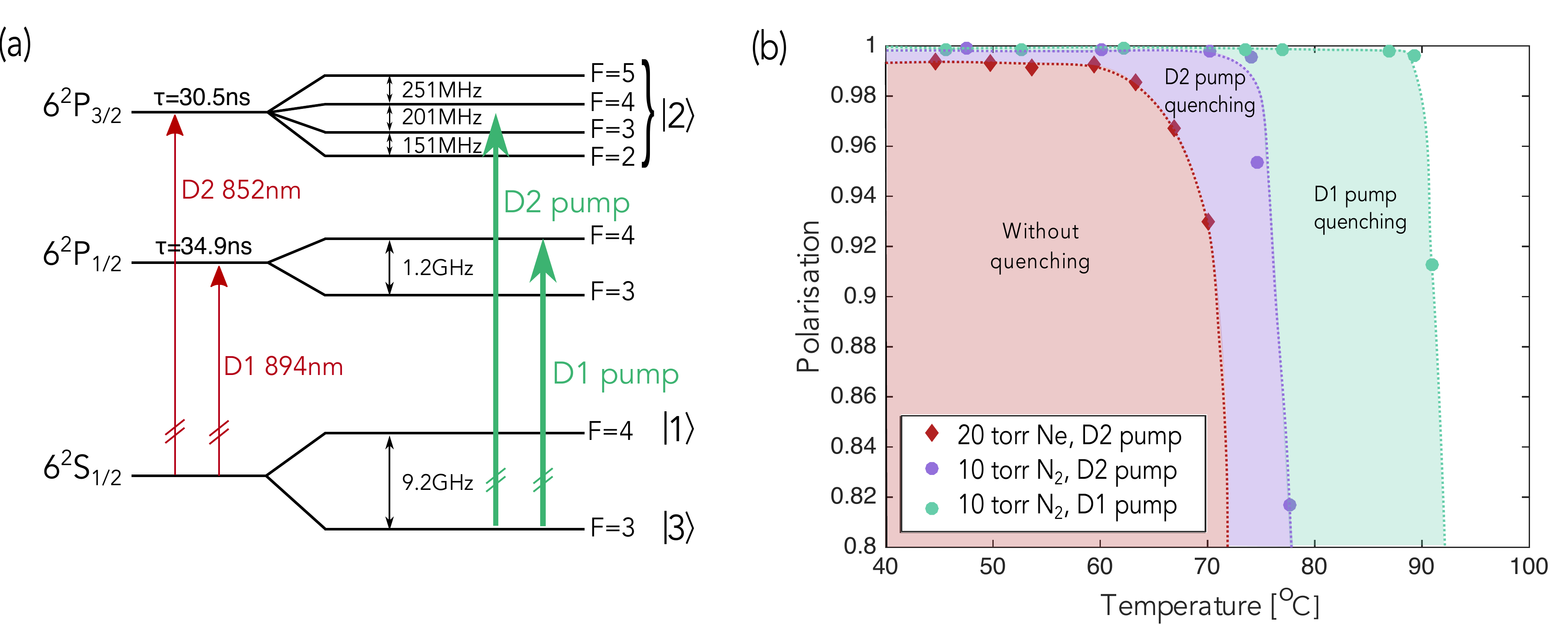}
\end{center}
\caption{(a) The energy levels of Cs, including hyperfine structure. The hyperfine levels in the ground state, $6^2\mathrm{S}_{1/2} F=3$ and $6^2\mathrm{S}_{1/2} F=4$ correspond to $\ket{3}$ and $\ket{1}$ respectively in Figure~\ref{figure1}(a). The two lowest excited states, $6^2\mathrm{P}_{1/2}$ and $6^2\mathrm{P}_{3/2}$, have lifetimes of 34.9ns and 30.5ns respectively. The hyperfine states of $6^2\mathrm{P}_{3/2}$ are not resolved and we use the entire manifold as state $\ket{2}$. We can optically pump the ensemble into $\ket{1}$ on either the D1 or D2 transition. All numerical values are taken from~\cite{Steck}, and the figure is not to scale.
(b) Measured spin-polarisation of Cs in different buffer gases. The lines are to guide the eye and are not a fit to the data.}
\label{fig:PolarisationData} 
\end{figure}

The spin-polarisation of Cs vapour as a function of temperature is shown in Fig.~\ref{fig:PolarisationData}(b). A description of how the spin-polarisation of a vapour is measured is given in the supplementary information. We see high spin-polarisation of the ensemble at low temperatures in all buffer gases, and then a sharp decline with increasing temperature which is due to radiation trapping. However, for temperatures above $70^\circ$C, the spin-polarisation is significantly higher in N$_2$ than in Ne due to the reduced radiation trapping in the molecular buffer gas. This demonstrates that collisional quenching is an effective way to enable spin-polarisation at very high temperatures.

We investigated optical pumping in N$_2$ buffer gas on different atomic transitions in Cs (the energy level structure is shown in~\ref{fig:PolarisationData}(a)). The quenching cross-section, which quantifies the probability of a quenching collision occurring, is $\sigma_Q=77\mathrm{\mathring{A}^{2}}$ for the D1 line and $\sigma_Q=69\mathrm{\mathring{A}^{2}}$ for the D2 line \cite{McGillis1968}. Furthermore, the D1 excited state has a longer lifetime (34.9ns compared to 30.5ns) and therefore there is more time for a collision to occur while the atom is in the excited state. Quenching is therefore more efficient on the D1 transition as the rate of collisional quenching is higher. 

Figure~\ref{fig:PolarisationData}(b) shows significantly higher spin-polarisation at temperatures above $75^\circ$C when the optical pumping light is tuned to the D1 atomic transition due to more efficient quenching. With 10~Torr of N$_2$ and by optically pumping on the D1 line we can now achieve spin-polarisations of over $99.5\%$ up to $90^{\circ}$C. Finally, we extract from our spin-polarisation measurements the optical depth, $d$, of the Cs vapour. Increasing the temperature from 70$^\circ$C to $90^\circ$C equates to an increase of $d$ from $5 \times 10^4$ to $2 \times 10^5$. This increase of a factor of 4 is shown to significantly enhance Raman interactions.\\
%

\section{Raman Memory}

\begin{figure*}[t!]
\begin{center}
\includegraphics[width=.9\linewidth]{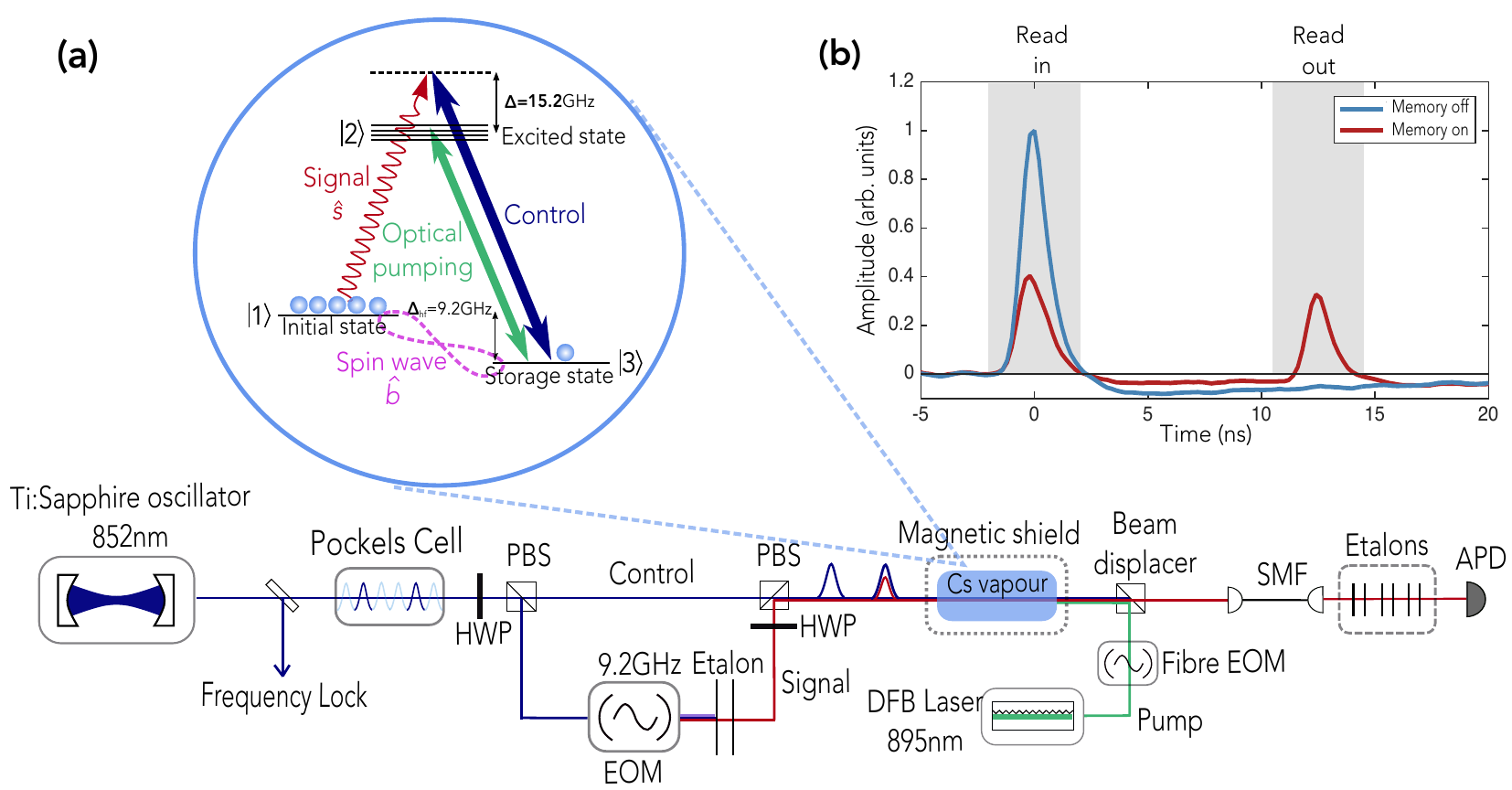}
\end{center}
\caption{(a) A schematic for the experimental setup of the bulk Raman memory. See the main text for more details. (b) An example recorded signal after the memory with (blue) the control field off (incident signal only, memory off) and (red) the control field on (memory on), with the background level subtracted. The shaded regions refer to the integration windows for the incident and retrieved pulses. }
\label{fig:MemoryData} 
\end{figure*}

The Raman memory protocol uses a $\Lambda$-level scheme to adiabatically transfer population between two ground states via an off-resonant stimulated Raman transition. We implement the memory in an ensemble of Cs atoms in the vapour phase, where the hyperfine states $6^2$S$_{1/2}$ $F=3$ and $F=4$ act as the storage state $\ket{3}$ and initial state $\ket{1}$, separated by $\Delta_\mathrm{hf}=9.2$~GHz, and the hyperfine manifold $6^2$P$_{3/2}$ is the excited state $\ket{2}$, as shown in Figure~\ref{fig:MemoryData}(a)~\cite{Reim2011}. The first step of the memory protocol is to spin-polarise the ensemble into the $F=4$ ground state, $\ket{1}$. Imperfect spin-polarisation decreases the memory efficiency due to the lower optical depth of the memory interaction optical transition which reduces the Raman interaction strength, and can also lead to adverse noise processes, such as spontaneous Raman noise, that will decrease the fidelity of the memory operation~\cite{Saunders2015}.

The Raman memory interaction is mediated by a strong control pulse which drives a two-photon Raman transition from $\ket{1} \rightarrow \ket{3}$. This adiabatically maps the signal mode $\hat{s}$ onto an excitation of the ground-state coherence of the ensemble, or spin-wave mode, $\hat{b}$. The interaction is described by a beam-splitter Hamiltonian 
\beq
\mathcal{H}_\mathrm{s} \propto C_\mathrm{s}\hat{s}\hat{b}^\dagger + \mathrm{h.c.}, \label{eq:BSH}
\eeq
with the coupling constant $C_\mathrm{s} \propto \sqrt{d\gamma\delta}\Omega/\Delta$, where $d$ is the resonant optical depth and $\gamma$ is the linewidth of the $\ket{1} \leftrightarrow \ket{2}$ transition, $\Omega$ is the control field Rabi frequency, $\delta$ is the control field bandwidth, $\Delta$ is the detuning from two-photon resonance \cite{JoshThesis}, and $\mathrm{h.c.}$ is the Hermitian conjugate. The signal is read out of the memory by applying a second control pulse which drives the reverse process, and this is a second beam splitter interaction with the same coupling strength $C_\mathrm{s}$. The total memory efficiency $\eta$ is proportional to $C_\mathrm{s}^4$, and hence scales with the square of the optical depth, $d$. Memory efficiencies of 21\% for single photon input states and 29\% for coherent states have been achieved with GHz bandwidths and $\mu$s storage times~\cite{Michelberger2015}. Here a neon buffer gas was used, and the Cs vapour temperature was 70$^\circ$C.

As well as the beam splitter interaction, this system also supports four-wave mixing (FWM) due to the control field coupling to state $\ket{1}$ and driving spontaneous anti-Stokes scattering. This interaction is described by a two-mode-squeezing Hamiltonian 
\beq
\mathcal{H}_\mathrm{a} \propto C_\mathrm{a}\hat{a}\hat{b} + \mathrm{h.c.}, \label{eq:TMSH}
\eeq
which produces excitations in both the anti-Stokes mode, $\hat{a}$, and the spin-wave, $\hat{b}$, followed by the beam-splitter Hamiltonian. The anti-Stokes coupling constant is given by $C_\mathrm{a}=C_\mathrm{s} \Delta/\Delta_\mathrm{a}$, where the detuning of this interaction is $\Delta_\mathrm{a}=\Delta+\Delta_\mathrm{hf}$. In the far-off-resonant limit, $\Delta \gg \Delta_\mathrm{hf}$, the magnitudes of the coupling strengths $C_\mathrm{s}$ and $C_\mathrm{a}$ are comparable. Four-wave mixing introduces gain into the system and can significantly enhance the amplitude of the retrieved pulse. Gain is always accompanied by noise~\cite{Sangouard2010} and we recently showed how this noise degrades the quantum features of single photons~\cite{Michelberger2015}, and how this can be suppressed for quantum light storage by operating the memory inside a low-finesse cavity~\cite{Saunders2015}. However, operating the memory inside a cavity is much more complex experimentally and we use a non-cavity memory system as a test bed for investigating the high coupling regime, even though it is not suitable for single-photon storage. 

We used the same memory setup as in previous demonstrations of the Raman memory~\cite{Michelberger2015} but with a higher operation temperature of up to $92^\circ\mathrm{C}$, which should enable significantly higher storage and retrieval efficiencies. The control and signal field pulses are both derived from a mode-locked, frequency-stabilised pulsed Titanium-Sapphire (Ti:Sa) laser which generates pulses with a central wavelength of $\sim$ 852nm and a spectral bandwidth of 1.2~GHz at a repetition rate of 80~MHz. Its exact central frequency is locked to be 15.2~GHz blue-detuned from the $6^2$S$_{1/2}$ ($F=3$) $\rightarrow$ $6^2\mathrm{P}_{3/2}$ transition. A Pockels cell is used to select the read-in and read-out pulses from this pulse train with a controllable separation time. The extinction ratio of the picked pulses after coupling into a single-mode fibre is approximately 40000:1. The beam is separated on a polarising beam splitter (PBS) and one arm passes through an electro-optic modulator (EOM) which is modulated at a frequency of 9.2~GHz to generate sidebands at $\pm$9.2~GHz, one of which is used as the input signal for the memory. An etalon is used to filter the carrier frequency and the blue sideband leaving only the red modulated sideband at the required signal frequency. The input coherent state amplitude is $|\alpha|^2\sim 10^ 5$. The signal and control fields are overlapped spatially and temporally and focused to a beam waist of $140\mu m$ at the centre of a vapour cell. The vapour cell is inside a $\mu$-metal shield to reduce the residual magnetic field by a factor of $\sim10^3$ and limit the dephasing of the ground-state coherence.  

 \begin{figure*}[t]
 \begin{center}
 \includegraphics[width=.9\linewidth]{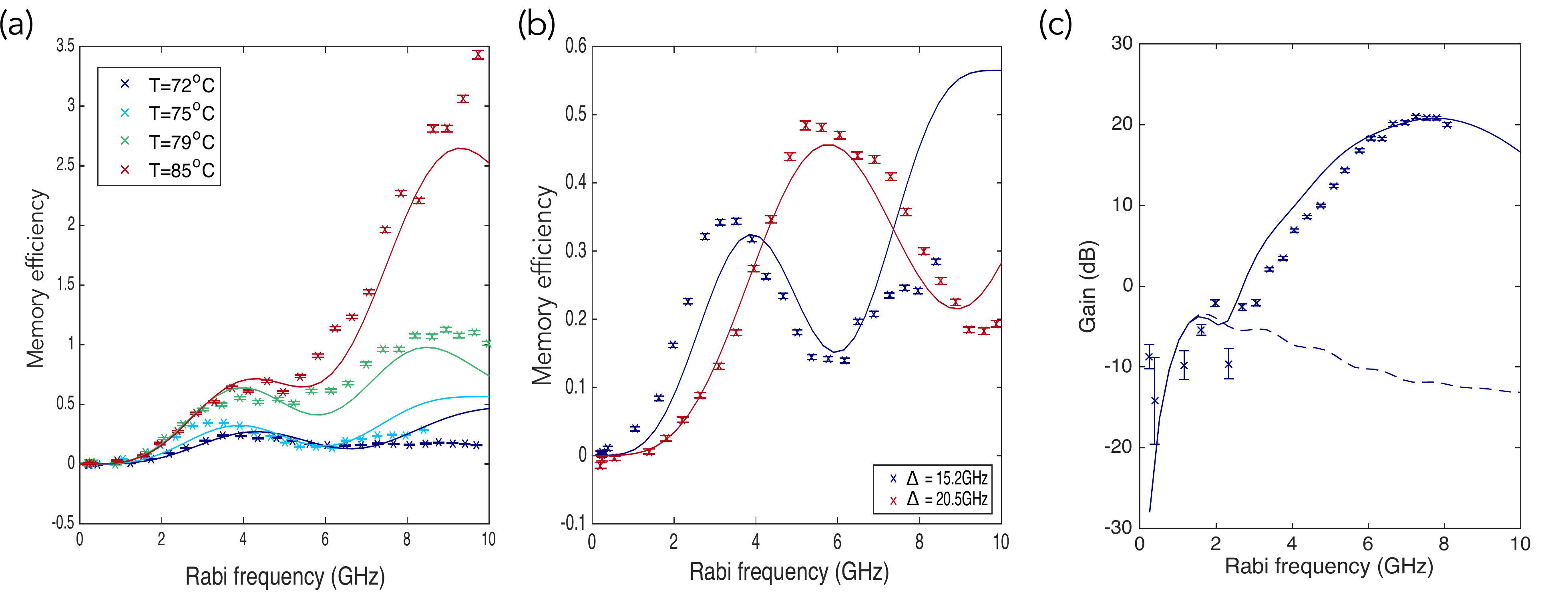}
 \end{center}
 \caption{(a) Memory efficiency as a function of the control Rabi frequency for different temperatures of the Cs vapour, with detuning fixed at $\Delta=15.2$~GHz. (b) Memory efficiency as a function of the control field Rabi frequency for two different detunings. The temperature of the Cs for these data is $75.2^\circ$C.(c) Memory efficiency in units of gain as a function of control field Rabi frequency for Cs vapour temperature of $\sim 92 ^\circ$C and detuning $\Delta = 15.2$~GHz. The dotted line shows the predicted memory efficiency when four-wave mixing is suppressed.}
 \label{AllData}
 \end{figure*}

We define the memory efficiency as the ratio of the pulse areas of the retrieved pulse and the input pulse. We measured the memory efficiency as a function of the control field Rabi frequency for different temperatures of the Cs ensemble, as shown in Figure~\ref{AllData}(a). For small control Rabi frequencies the memory efficiency increases quadratically with $\Omega$, since the Raman coupling strength $C_\mathrm{s}$ is proportional to $\sqrt{d}\frac{\Omega}{\Delta}$. However, as $\Omega$ increases, the strong control field induces an AC Stark shift between states $\ket{2}$ and $\ket{3}$, which perturbs the two-photon resonance condition and reduces the memory efficiency~\cite{JoshThesis}. For each temperature this occurs at $\Omega \approx 4 $~GHz for $\Delta = 15.2$~GHz, where the AC Stark shift becomes comparable to the spectral bandwidth of the pulse, and we see a decrease in memory efficiency at this point. However, at even higher Rabi frequencies we see that the memory efficiency increases dramatically, even above 100\%, due to the FWM gain process. The average number of anti-Stokes photons produced is proportional to $\sinh^2(C_\mathrm{a}/\delta)$, where $\delta$ is the pulse bandwidth, and therefore at high control field Rabi frequency, the four-wave mixing process increases approximately exponentially in $\Omega$ and this gain interaction dominates and amplifies the input signal. We can see in Figure~\ref{AllData}(a) that the measured memory efficiency is significantly greater at higher temperatures. This increase is a direct consequence of the higher Raman coupling strength due to increased optical depth which was enabled by improving the spin-polarisation at higher temperatures. 

We investigated operating the Raman memory further from resonance to reduce linear absorption of the signal and decrease the AC Stark shift, as shown in Figure~\ref{AllData}(b). The Raman interaction strength is proportional to $1/\Delta^2$ and therefore for small Rabi frequencies we see a higher memory efficiency when the detuning from resonance is smaller. However, the AC Stark shift $\Delta E$ is proportional to $\Omega^2/\Delta$, and therefore the memory efficiency reaches higher values when we operate further from resonance due to the suppression of the AC Stark shift. This shows an additional benefit of the increased Raman interaction strength: the increase in optical depth means that we can operate further from resonance while still maintaining strong Raman coupling and therefore mitigate the AC Stark effect and achieve higher memory efficiency.
 
To model these data we numerically solve the linearised Maxwell-Bloch equations in one dimension in the adiabatic limit~\cite{Michelberger2015}. This model assumes that the excited state can be adiabatically eliminated and that the population of the states remains constant throughout the interaction, as the size of the signal field is much smaller than the number of atoms in the ensemble. The model includes four-wave mixing gain and the AC Stark shift, and Figure~\ref{AllData} shows that it captures all features of the data. 

The data in Figure~\ref{AllData}(c) show the memory efficiency when the ensemble temperature is $92^\circ$C, and the storage efficiency increases up to $\gtrsim12000\%$. This demonstrates extremely strong amplification of the input signal of over 20dB due to the strong Raman interaction. This technique therefore has potential applications as an amplified delay-line for optical signals in a room temperature system for optical signal processing applications. These results could also be applied to the generation of strongly-squeezed light via four-wave mixing in alkali vapours~\cite{Lett2008}. Adding a molecular buffer gas to increase the optical depth of a spin-polarised alkali vapour could enhance the Raman interaction strength and enable even higher two-mode squeezing.

%
Furthermore, these results can be applied to a cavity-enhanced Raman memory protocol where four-wave mixing noise can be effectively suppressed to allow low-noise storage of single photons~\cite{Novikova2016,Saunders2015}. The dashed line in Figure~\ref{AllData}(c) shows the predicted memory efficiency if there is no four-wave mixing, and this gives an indication of the potential performance of this memory for quantum-level storage. We predict that the memory efficiency in the cavity-enhanced Raman memory presented in~\cite{Saunders2015} would increase from 9.5\% to 18.8\% with the addition of a molecular buffer gas. The strong Raman coupling enabled by collisional quenching enhances the memory interaction strength and opens the route to achieving high-efficiency storage of single photons in the near future.

\section*{Acknowledgements}

This work was supported by the UK Engineering and Physical Sciences Research Council through Standard Grant No. EP/J000051/1, Programme Grant No. EP/K034480/1, and the EPSRC NQIT Quantum Technology Hub. We acknowledge support from the Air Force Office of Scientific Research: European Office of Aerospace Research and Development (AFOSR EOARD Grant No. FA8655-09-1-3020). JN acknowledges a Royal Society University Research Fellowship, and DJS acknowledges an EU Marie-Curie Fellowship No. PIIF-GA-2013-629229. P.M.L. acknowledges a European Union Horizon 2020 Research and Innovation Framework Programme Marie Curie individual fellowship, Grant Agreement No. 705278, and B. B. acknowledges an H2020 Future and Emerging Technologies Grant QCUMBER (Grant Award number 665148). I.A.W. acknowledges an ERC Advanced Grant (MOQUACINO). C.Q. was supported by the China Scholarship Council (CSC Grant No. 201406140039). S.E.T. and J.H.D.M are supported by EPSRC via the Controlled Quantum Dynamics CDT under Grants EP/G037043/1 and EP/L016524/1. 

\section*{Author contributions}
This experiment was conceived and designed by D.J.S., and was performed by S.E.T. with assistance from D.J.S., P.M.L., B.B., C.Q. and A.F. The optical pumping data were analysed by S.E.T. with assistance from J.H.D.M. The diffusion measurements were performed by J.H.D.M. and K.T.K. and analysed by J.H.D.M. The Raman memory data was modelled theoretically by J.N. This project was supervised by D.J.S., J.N. and I.A.W.

\section*{Competing financial interests}
The authors declare no competing financial interests.

\bibliography{BulkMemoryBib}
\bibliographystyle{unsrt}

\clearpage 
\onecolumngrid

\setcounter{figure}{0}

\renewcommand{\thefigure}{S\arabic{figure}}

\begin{center}
\large{\bf Supplementary Material}
\end{center}

\section*{Spin Polarisation Measurements} 

\begin{figure}[h]
\begin{center}
\includegraphics[width=0.5\linewidth]{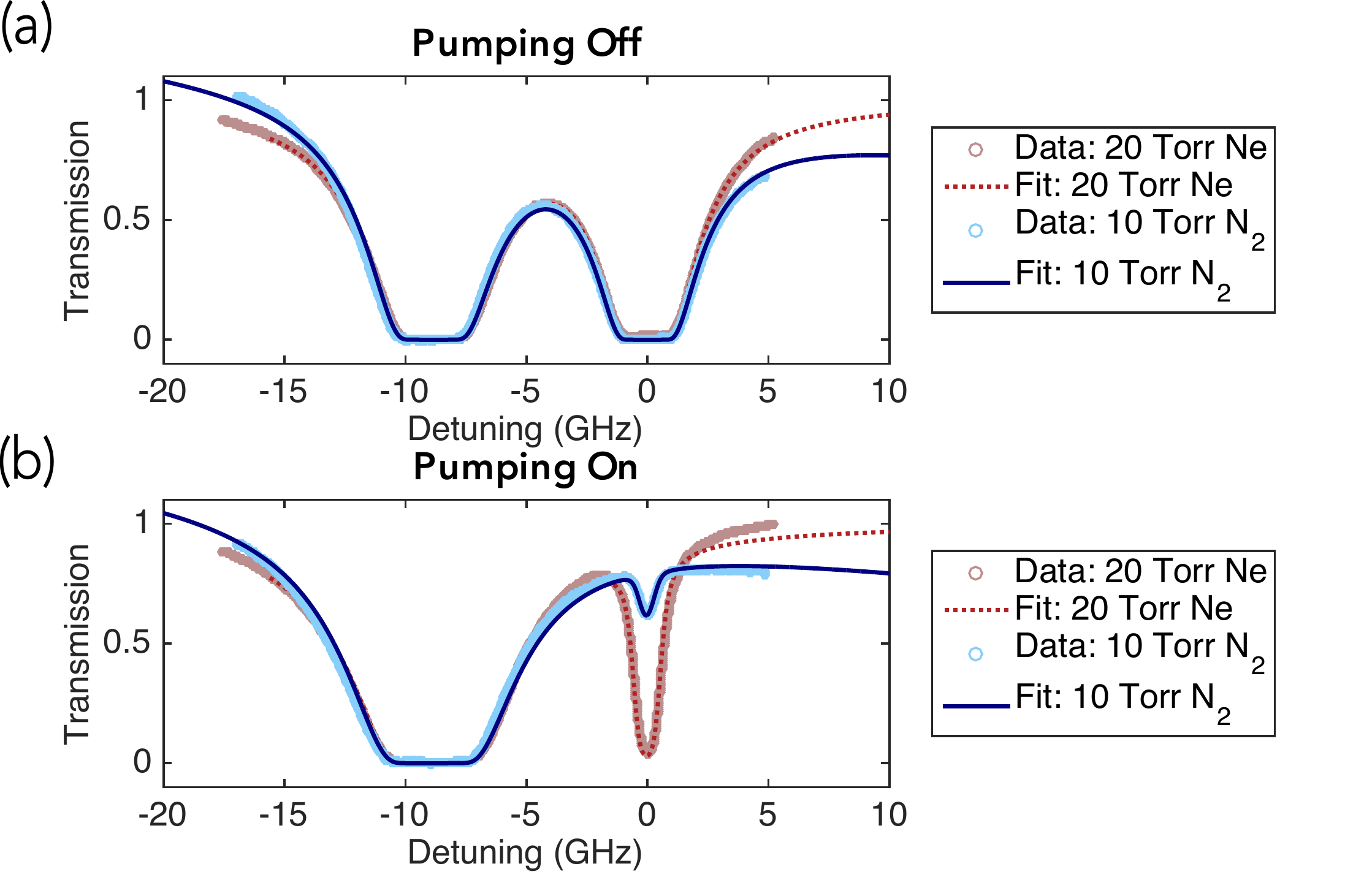}
\end{center}
\caption{ Spin-Polarisation measurements. The transmission of a probe beam through vapour cells containing Cs and a buffer gas of either 10 Torr of N$_2$ or 20 torr of Ne is recorded as it scans in frequency across the D2 transitions. The temperature for these data sets is $\sim 70^\circ$C. When the pumping is off (a) we see comparable absorption from the two ground states as they are equally populated in thermal equilibrium. When the pump beam is turned on (b) we see a significant transfer of population from the $F=3$ to $F=4$ ground state. The spin-polarisation, extracted from fitting the absorption profile, is 96.7\% for 20 Torr of Ne, and for 10 Torr of N$_2$ it is 99.9\%.}
\label{fig:PolMeasurements} 
\end{figure}

We measured the spin-polarisation, $P$, of Cs vapour as a function of temperature with 20 Torr of Ne or 10 Torr of N$_2$ buffer gas. We recorded the transmission of a weak probe beam through the vapour as it scans in frequency across the D2 line (the atomic line structure of Cs is shown in Figure~\ref{fig:PolarisationData}(a)). When the optical pumping light is blocked we see the two pressure- and Doppler-broadened absorption profiles corresponding to the transitions from the $F=3$ $(\ket{3})$ and $F=4$ $(\ket{1})$ ground states, as shown in Fig.~\ref{fig:PolMeasurements}(a). When the optical pumping beam is turned on we see significantly less absorption from transitions from the $F=3$ state (Fig.~\ref{fig:PolMeasurements}(b)) which demonstrates the transfer of population from $F=3\rightarrow F=4$ due to the optical pumping.  The frequency scan is calibrated by also passing the probe through a room temperature Cs cell with no buffer gas in, in a saturated absorption spectroscopy setup to observe the Doppler-free absorption profile. The overall slope of the data is due to the feed forward mechanism of the ECDL probe beam, and is taken into account in the fit.

The absorption and dispersion of alkali vapours is well studied, for example in~\cite{Siddons2009,Munns2016}, and we can calculate the transmission of a probe beam by summing over the Voigt profile of each atomic transition. We fit the data numerically and extract the optical depth, $d$, of the vapour and the ratio of atoms in the two ground states, and Figure~\ref{fig:PolMeasurements} shows a good fit to the data. The temperature of the vapour is also extracted from the Doppler- and pressure-broadening widths of the fit. \\


\section*{Measurements of atomic diffusion} 

In order to investigate the optimal buffer gas species and pressure, the diffusion coefficients, $D$, of Cs in the presence of Ne and N$_2$ were measured at room temperature. We follow a method described in~\cite{Parniak2013}, where a small central region of atoms in a vapour are optically pumped, and then we probe the entire vapour. The optically pumped atoms are transparent to the probe beam and therefore give rise to a region of decreased optical depth. We observe how the spatial distribution of this region of decreased optical depth varies over time and therefore can obtain information about the rate of diffusion of Cs atoms.

The pump and probe beam are derived from the same diode which is locked to the D2 absorption line and switched on and off using an acousto-optic modulator. The beams are orthogonally polarised and counterpropagating to separate the intense pump and weak probe light. The probe pulse is short so as to not dynamically affect the populations whilst imaging. 

Due to the spatial asymmetries of the optical depth profile, a simple Gaussian approximation of the radial distribution is not sufficient and we consider the more general approach to analyse the diffusion of the pumped atoms presented in \citep{Parniak2013}. 

The general diffusion equation is given by
\beq
\frac{\partial \phi(\mathbf{r},t)}{\partial t}=D\nabla^2 \phi(\mathbf{r},t)
\eeq
for a spatially and temporally constant diffusion coefficient, $D$.  

Transforming to the spatial Fourier domain in the transverse coordinate \mbox{$\rho=x\hat{\mathbf{e}}_x+y\hat{\mathbf{e}}_y$}
\beq
\mathcal{F}\left\{\Delta d(\rho,t)\right\}=\frac{1}{2\pi}\iint\ \mathrm{d} \rho\,\Delta d(\rho,t)\mathrm{e}^{-\mathrm{i}\mathbf{k}_\perp\cdot\rho}\nonumber\\
\eeq
for which the solution of the diffusion equation gives
\begin{equation*}
\begin{aligned}
\mathcal{F}\left\{\Delta d(\rho,t)\right\}
=\mathcal{F}\left\{\Delta d(\rho,0)\right\}\,\mathrm{e}^{-\left[\gamma_0+D\vert \mathbf{k}_\perp\vert^2\right]t}\\
=\mathcal{F}\left\{\Delta d(\rho,0)\right\}\,\mathrm{e}^{-\gamma(k_\perp)t}
\end{aligned}
\end{equation*}

where $\gamma_0$ is explicitly added to account for additional relaxation mechanisms.  The interpretation of the above is that components of the perturbed optical depth with a given ``spatial frequency'' decay exponentially with rate \mbox{$\gamma_0+Dk_\perp^2$}.

For a given pressure of each buffer gas, we therefore fit an exponential decay to spatial frequency components corresponding to lengthscales between the pixel size ($k_\mathrm{max}\sim$\mbox{$(1/0.0259)\,\mathrm{mm}^{-1}$}) and the probe beam size ($k_\mathrm{min}\sim$\mbox{$(1/10)\,\mathrm{mm}^{-1}$}).  The decay constant, $\gamma(k_\perp)$, of the amplitude of each element in the Fourier transformed image with a given wavenumber \mbox{$k_\perp=\sqrt{k_x^2+k_y^2}$} is obtained by fitting the time dependence for a given element to a decaying exponential.  The extracted values are then fitted to a quadratic \mbox{$\gamma=\gamma_0+Dk_\perp^2$} to yield the diffusion constant $D$ for each buffer gas at a given pressure.  

\begin{figure}
 \begin{center}
 \includegraphics[width=.9\linewidth]{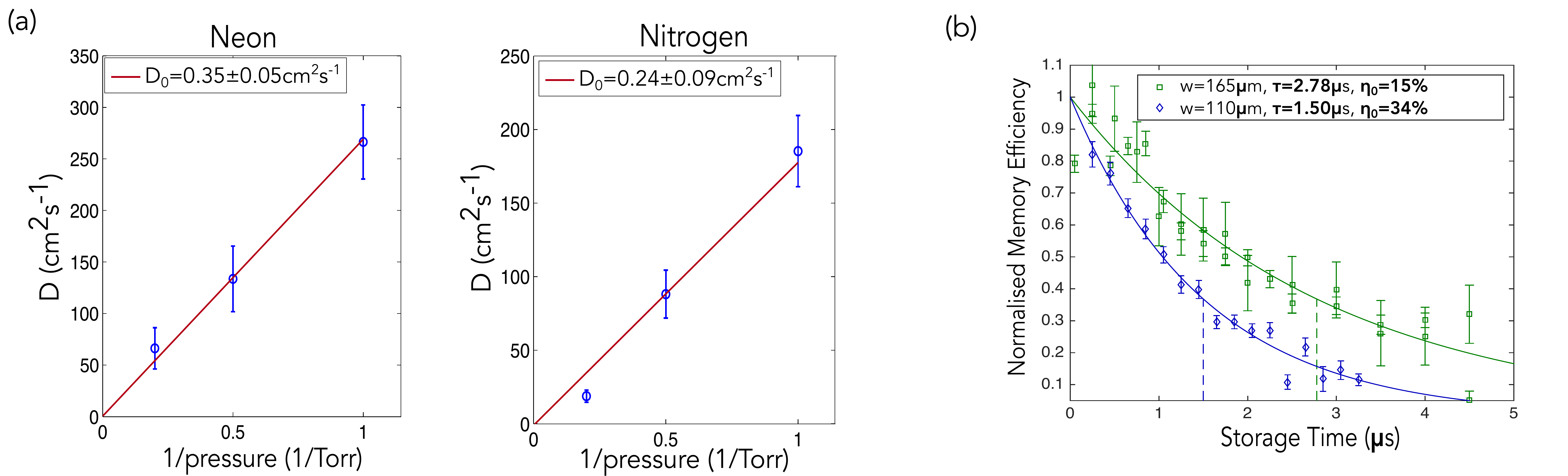}
 \end{center}
 \caption{(a) Extraction of $D_0$ for Cs in Ne and N$_2$ buffer gases. (b) Memory efficiency as a function of storage time for different waists, $w$, of the signal and control fields, normalised to the memory efficiency for a storage time of 12.5ns ($\eta_0$). An exponential is fitted to the data as this is the decay we expect for diffusion-limited lifetime, and the lifetime, $\tau$, is given as the 1/e point. }
 \label{DiffusionResults}
 \end{figure}

The diffusion properties in the presence of a given buffer gas can be characterised by the representative parameter $D_0$ \citep{Happer1972} which is related to the diffusion constant $D$ at a given pressure, $p$, by:
\beq
D_0=\frac{p}{760\,\mathrm{Torr}}D(p)\label{eqn_HapperD0}
\eeq
The extracted $D_0$ parameters using Eq. (\ref{eqn_HapperD0}) are plotted in Fig.~\ref{DiffusionResults}(a) and given in Table~\ref{tab_D0vals} in the main text. The error bars are generated by performing the automated fitting routine independently over each quadrant of the image, and the fits have been constrained so as to pass through zero diffusion at ``infinite pressure''. \\

\section*{Memory Lifetime} We investigate the lifetime of the memory by measuring the memory efficiency as a function of storage time. The lifetime of the memory is limited by two main factors: diffusion of the Cs atoms and magnetic dephasing of the ground state coherence. If the lifetime is diffusion-limited then an increase in beam waist will increase the interaction time and therefore increase the memory lifetime. However, if the lifetime is instead limited by magnetic dephasing there will be negligible dependence of the lifetime of the size of the beams.

The results, shown in Fig.~\ref{DiffusionResults}(b) and show a significantly different lifetime from 1.5$\mu$s for a waist of 110$\mu$m to 2.8$\mu$s for a waist of 165$\mu$m. This indicates that the dominant dephasing mechanism is diffusion of the Cs atoms out of the beam. The magnetic shielding used in the experiment, three concentric cylinders of $\mu$-metal, was therefore sufficient to increase the magnetic dephasing timescale to more than 1$\mu$s. The memory efficiency decreases if we increase the beam waist as the control Rabi frequency is lower and therefore there is a reduced interaction strength. 


\end{document}